\documentclass[fleqn,11pt]{article}
\usepackage{amsmath,amssymb,amsfonts}

\def\noi{\noindent}

\newcommand{\Title}[1]{\noi {{\Large\bf #1}}\\[1ex]}

\newcommand{\Author}[2]{\noi{\bf #1}\\[2ex]\noi{\normalsize\it #2}\\}

\newcommand{\Abstract}[1]{\vskip 2mm \begin{center}
        \parbox{16.4cm}{\small\noi #1} \end{center}\medskip}

\def\nqq{\hspace*{-2em}}
\def\nhq{\hspace*{-0.5em}}

\def\cm{\hspace*{1cm}}

\def\Jl#1#2{#1 {\bf #2},\ }

\def\ApJ#1 {\Jl{Astroph. J.}{#1}}
\def\CQG#1 {\Jl{Class. Quantum Grav.}{#1}}
\def\DAN#1 {\Jl{Dokl. AN SSSR}{#1}}
\def\GC#1 {\Jl{Grav. Cosmol.}{#1}}
\def\GRG#1 {\Jl{Gen. Rel. Grav.}{#1}}
\def\JETF#1 {\Jl{Zh. Eksp. Teor. Fiz.}{#1}}
\def\JETP#1 {\Jl{Sov. Phys. JETP}{#1}}
\def\JHEP#1 {\Jl{JHEP}{#1}}
\def\JMP#1 {\Jl{J. Math. Phys.}{#1}}
\def\NPB#1 {\Jl{Nucl. Phys. B}{#1}}
\def\NP#1 {\Jl{Nucl. Phys.}{#1}}
\def\PLA#1 {\Jl{Phys. Lett. A}{#1}}
\def\PLB#1 {\Jl{Phys. Lett. B}{#1}}
\def\PRD#1 {\Jl{Phys. Rev. D}{#1}}
\def\PRL#1 {\Jl{Phys. Rev. Lett.}{#1}}

\def\lal{&&\nqq {}}
\def\eq{Eq.\,}
\def\eqs{Eqs.\,}
\def\beq{\begin{equation}}
\def\eeq{\end{equation}}
\def\bear{\begin{eqnarray}}
\def\bearr{\begin{eqnarray} \lal}
\def\ear{\end{eqnarray}}
\def\earn{\nonumber \end{eqnarray}}

\def\nnn{\nonumber\\ \lal }

\def\yy{\\[5pt] {}}

\def\dst{\displaystyle}

\def\fracd#1#2{{\dst\frac{#1}{#2}}}

\def\Half{{\fracd{1}{2}}}

\def\const{{\rm const}}

\def\ep{\epsilon}

\def\eqn#1{\eq\eqref{#1}}
\def\rf{\eqref}
\def\mn{_{\mu\nu}}
\def\MN{^{\mu\nu}}

\def\cE{{\cal E}}

\def\kappa{\varkappa}
\def\ep{\epsilon}

\def\ssph{static, spherically symmetric}

\def\asflat{asymptotically flat} 

\begin{document}
\twocolumn[

\Title{Dyonic configurations in nonlinear electrodynamics\yy coupled to general relativity}
	
\Author{K. A. Bronnikov}
       {VNIIMS, Ozyornaya ul. 46, Moscow 119361, Russia;\\
        Institute of Gravitation and Cosmology, Peoples' Friendship University of Russia 
         (RUDN University),\\ \hspace*{1em}  ul. Miklukho-Maklaya 6, Moscow 117198, Russia;\\
	National Research Nuclear University ``MEPhI'', Kashirskoe sh. 31, Moscow 115409, Russia}

\Abstract   
 {We consider static, spherically symmetric configurations in general relativity, supported by  
  nonlinear electromagnetic fields with gauge-invariant Lagrangians depending on the single 
  invariant $f = F_{\mu\nu} F^{\mu\nu}$. After a brief review on black hole (BH) and 
  solitonic solutions, obtained so far with pure electric or magnetic fields, an attempt is made  
  to obtain  dyonic solutions, those with both electric and magnetic charges. A general scheme 
  is suggested, leading to solutions in quadratures for an arbitrary Lagrangian function $L(f)$ 
  (up to some monotonicity restrictions); such solutions are expressed in terms of $f$ 
  as a new radial coordinate instead of the usual coordinate $r$. For the truncated Born-Infeld 
  theory (depending on the invariant $f$ only), a general dyonic solution is obtained in terms of $r$. 
  A feature of interest in this solution is the existence of a special case with a self-dual 
  electromagnetic field, $f \equiv 0$ and the Reissner-Nordstr\"om metric.    
}

] 

\section{Introduction. A brief review}

  In the framework of general relativity (GR) and its extensions, nonlinear electrodynamics (NED) 
  as a possible material source of gravity attracts much attention since, among other reasons, 
  it leads to many space-time geometries of interest, in particular, regular black holes and starlike or
  solitonlike configurations.

  The history of NED apparently begins with Born and Infeld's effort to remove the central 
  singularity of a point charge by generalizing Maxwell's theory \cite{B-Inf}, 
  the version of NED of Heisenverg and Euler motivated by particle physics \cite{EuH}, and their 
  further extension by Plebanski \cite{Pleb} in the framework of special relativity. 

  In this paper, we consider \ssph\ configurations in NED coupled to GR and begin with a brief 
  overview of the results obtained thus far in this area. For such systems, one usually considers 
  the action
\beq            \label{S}
	S = \Half \int \sqrt {-g} d^4 x [R - L (f)], 	\ \ \   f = F\mn F\MN, 
\eeq
  with an arbitrary function $L (f)$  ($F\mn$ is the Maxwell tensor, and units with $c = 8\pi G =1$ are
   used). Then, assuming static spherical symmetry, the stress-energy tensor (SET) satisfies 
  the condition $T^t_t = T^r_r$, hence, due to the Einstein equations, the metric can be written as 
\beq            \label{ds} 
	ds^2 = A(r) dt^2 - \frac{dr^2}{A(r)} - r^2 (d\theta^2+\sin^2 \theta d\phi^2).
\eeq
  The only nonzero components of $F\mn$ are $F_{tr} =- F_{rt}$ (a radial electric field) 
  and $F_{\theta\phi} = - F_{\phi\theta}$ (a radial magnetic field). The Maxwell-like equations
  $\nabla_\mu (L_f F\MN) = 0$ and the Bianchi identities $\nabla_\mu {}^*F\MN = 0$ 
  for the dual field $^* F\MN$ lead to
\beq              \label{F_mn}
                      r^2 L_f F^{tr} = q_e, \cm    F_{\theta\phi} = q_m\sin\theta,
\eeq
  where $q_e$ and $q_m$ are the electric and magnetic charges, respectively, and $L_f \equiv dL/df$. 
  Accordingly, the nonzero SET components are 
\bearr    \nhq       \label{SET}
        T^t_t = T^r_r = \Half L + f_e L_f, \quad T^\theta_\theta = T^\phi_\phi = \Half  L - f_m L_f,
\nnn
	f_e = 2 F_{tr}F^{rt}=\frac{2q_e^2}{L_f^2 r^4}, \ \ \ 
	f_m = 2 F_{\theta\phi}F^{\theta\phi} = \frac {2q_m^2}{r^4},
\nnn
\ear
  thus the invariant $f$ is $f = f_m - f_e$. The metric function $A(r)$ is determined from the 
  Einstein equations as 
\beq          \label{A}
	A(r) = 1 -\frac{2M(r)}{r}, \ \ \ M(r) = \frac 12 \int \cE(r) r^2 dr,  
\eeq
  where $\cE(r) \equiv T^t_t$ is the energy density, and $M(r)$ is the mass function. It is a general 
  relation  \cite{k-NED}, but it is only a part of a possible solution: the latter requires knowledge 
  of $L(f)$ and both electric and magnetic fields.

  For configurations with an electric field only, the general solution was obtained in 1969 by Pellicer 
  and Torrence \cite{Pel-T}. They found, under some reasonable assumptions, that the solution can 
  have well-behaved electromagnetic and metric tensors at the center $r = 0$. However, a no-go
  theorem proved in \cite{B-Shi} showed that there is no such function $L(f)$ having a Maxwell
  weak-field limit ($L\sim f$ as $f\to 0$) that the electric solution described by
  \rf{ds}, \rf{F_mn}, \rf{A} has a regular center. The reason is that at such a center 
  the electric field should be zero but the field equations then imply $f L_f^2 \to \infty$, hence 
  $L_f \to \infty$ as $r\to 0$.

  In 1998 and 1999 appeared a few papers by Ayon-Beato and Garcia (see, e.g., \cite{AB-G}),   
  where some special cases of the Pellicer-Torrence solution were presented, 
  describing configurations with or without horizons (that is, BH or solitonic ones), having a regular 
  center and a Reissner-Nordstr\"om asymptotic behavior at large $r$. Each of these examples seemed 
  to violate the above no-go theorem. Trying to clarify the situation, I found the following explanation 
  \cite{B-comment}: in all such cases, in a ``regular'' solution there are different functions $L(F)$ at 
  large and small $r$: at large $r$ we have $L\sim f$ whereas at small $r$ the theory is strongly 
  non-Maxwell ($f\to 0$ but $L_f \to \infty$, in agreement with the no-go theorem). An inspection 
  showed that it is indeed the case in all examples \cite{k-NED}.

  It was further shown in \cite{k-NED} that a regular center is also impossible in dyonic 
  configurations, with both $q_e\ne 0$ and $q_m \ne 0$, if $L(f)\sim f$ as $f\to 0$. However, 
  purely magnetic regular configurations, both BH and solitonic ones, are possible and are readily 
  obtained under the condition $L(f) \to L_\infty < \infty$ as $f  \to \infty$ \cite{k-NED}.  
  Electric models with the same regular metrics can be obtained from the magnetic ones using the 
  so-called FP duality \cite{k-NED} (not to be confused with the familiar electric-magnetic duality 
  in Maxwell's theory) that connects solutions with the same metric corresponding to {\it different\/} 
  NED theories. However, unlike the magnetic solutions, the electric ones suffer serious problems 
  connected with multivaluedness of $L(f)$ and a singular behavior of the electromagnetic fields  
  on the corresponding branching surfaces \cite{k-NED}.

  Many results of interest were obtained since then.   

  Burinskii and Hildebrandt \cite{Bur1} showed that the above no-go theorem for electric solutions 
  may be circumvented by considering a kind of phase transition on a certain sphere, outside which 
  there is a purely electric field $F\mn$ but inside which it is purely magnetic. An external observer then 
  sees an electrically charged BH or soliton. They also pointed out links between these regular 
  GR/NED solutions and the superconductivity and confinement phenomena \cite{Bur1,Bur2}.  

  Some authors considered GR/NED solutions with NED Lagrangians depending on both 
  electromagnetic invariants $f = F\mn F\MN$ and ${\tilde f } 
  = \ep_{\mu\nu\rho\sigma}F\MN F^{\rho\sigma}$, where  $\ep_{\mu\nu\rho\sigma}$ is the 
  Levi-Civita unit completely antisymmetric tensor \cite{Diaz-10,Ruf-13,Sche-15}. It should be  
  mentioned that the most famous versions of NED, the Born-Infeld \cite{B-Inf}
  and Euler-Heisenberg \cite{EuH} theories belong to this class.

  A straightforward extension of \ssph\ NED solutions to GR with a nonzero cosmological constant
  $\Lambda$, leading to their (anti-)de Sitter (dS and AdS) asymptotic behavior, was considered, 
  in particular, in \cite{Mat-09, Mat-13, Fer-15}. If we add $-2\Lambda$ to $R$ in the action 
  \rf{S}, the only change in the expression \rf{A} for the metric is that the term $ - \Lambda r^2/3$ 
  is added to $A(r)$. 

  Thermodynamic properties of regular NED BHs were discussed in 
  \cite{Bret-05, Yun-08, Bret-14, Cul-14, Ma-15, Bal-15, Kru-16, FW-16}, in particular, the first law 
  of thermodynamics, heat capacity, the validity of Smarr's formula for the BH mass, etc. 

  Cylindrically symmetric counterparts of the spherical solutions were considered 
  in \cite{we-02}, with different directions of the electric or magnetic fields: radial, 
  longitudinal and azimuthal ones. There are no BH solutions, while the field behavior on a possible regular
  axis (especially in the case of radial fields) is somewhat similar to that at a regular center in spherical
  symmetry, and special conditions appear if asymptotic flatness is required. 
  
  More complicated axially symmetric regular GR/NED configurations were discussed in
  \cite{Bam-13, Dym-15a, Dym-15b}. It was claimed, in particular, that in these models the Kerr ring 
  singularity is replaced by a ring with superconducting current that serves as a nondissipative source 
  of the asymptotically Kerr-Newman geometry \cite{Dym-15b}. It seems that this area of utmost 
  interest is only beginning to be studied. 
  
  It has been shown that NED can be a source of evolving wormholes \cite{Arel-06, Boe-07, Arel-09},
  while static wormholes are impossible since the SET \rf{SET} (though marginally) satisfies the 
  weak and null energy conditions. 

  Many authors studied the stability properties of NED BHs 
  \cite{Mor-03, Dym-05, Bret-05s, Jin-13, Jin-14, Ma-15}
  and their quasinormal modes related to different kinds of perturbations
  \cite{Lem-13, Jin-13, Jin-14, Tosh-15, Fer-15} in cases where these BHs are stable. 
  In particular, simple conditions on the NED Lagrangian have been derived, which imply linear
  stability in the domain of outer communication \cite{Mor-03}, and linear stability has been verified 
  for a number of particular examples.
  
  Quantum effects in the fields of NED BHs were studied in \cite{Mat-02, Mat-13}.

  This list is certainly incomplete: it does not include numerous papers devoted to studies of special 
  cases of electric and magnetic solutions, to say nothing of such subjects as gravitational lensing, 
  particle motion and matter accretion in the fields of NED BHs, their counterparts in scalar-tensor, 
  $f(R)$ and multidimensional theories of gravity, inclusion of dilaton-like interactions, non-Abelian fields, 
  constructions with thin shells, etc.

  A substantial gap in these studies is the absence of dyonic solutions (at least to our knowledge), and 
  we here try to consider this challenging problem. Before that, for completeness and comparison, 
  we recall the methods of obtaining pure magnetic and electric solutions.  

\section{Pure magnetic and electric solutions} 

  Pure magnetic solutions ($q_e = 0$) are obtained from \eqs \rf{SET} and \rf{A} quite easily.
  Indeed, if $L(f)$ is specified, then, since now $f = 2q_m/r^4$, the function $\cE(r) = L/2$ is known 
  from \rf{SET}, and the metric function $A(r)$ is found by integration in \rf{A}. If, on the contrary, 
  $A(r)$ is known (or chosen at will), then $\cE(r) = L(f)/2$ is found from \rf{A}, and $L(f)$ is
  restored since $f = 2q_m/r^4$. Solutions with a regular center are obtained if we require 
  $A(r) = 1 + O(r^2)$ at small $r$ (and this in turn requires $L \to L_0 < \infty$ as $f\to\infty$ 
  \cite{k-NED}), \asflat\ configurations require $A(r) = 1 - 2m/r + o(1/r)$, where $m$ is the 
  Schwarzschild mass, and, in the case $\Lambda \ne 0$, asymptotically (A)dS solutions are 
  obtained by adding $ - \Lambda r^2/3$ to $A(r)$ in \rf{A}. This is an easy way of constructing 
  regular magnetic BH and solitonic solutions, used by many authors, probably beginning with 
  \cite{k-NED}.     

  Pure electric solutions ($q_e\ne 0$, $q_m=0$) are obtained in a similar way by using a 
  Hamiltonian form of NED \cite{Pel-T, AB-G}, obtained from the original, Lagrangian one 
  by a Legendre transformation: one introduces the tensor $P\mn = L_f F\mn$ with its invariant 
  $p = -P\mn P\MN$ and considers the Hamiltonian-like quantity $H = 2f L_f  - L = 2T^t_t$ as 
  a function of $p$; then $H(p)$ can be used to specify the whole theory. One has then
\beq
          L = 2p H_p - H, \quad  L_f H_p = 1, \quad  f = p H_p^2.
\eeq
  with $H_p \equiv dH/dp$. Thus we have simply $p= 2q_e^2/r^4$, and specifying $H(p) = 2\cE(r)$, 
  we directly find $M(r)$ and $A(r)$ from \rf{A}. On the contrary, specifying $A(r)$, 
  from \rf{A} we determine $\cE(r) = H(p)/2$.  

  A regular center $r=0$ requires a finite limit of $H$ as $p \to \infty$, then the integral in \rf{A}  
  gives the mass function for given $q_e$ that provides regularity. However, in any regular 
  \asflat\ (or (A)dS) solution $f = 0$ at both $r=0$ and $r=\infty$, so $f$ inevitably has at least 
  one maximum at some $p=p^*$, violating the monotonicity of $f(p)$, which is necessary for 
  equivalence of the $f$ and $p$ frameworks. It has been shown \cite{k-NED} that at an 
  extremum of $f(p)$ the Lagrangian function $L(f)$ suffers branching, its plot forming a 
  cusp, and different functions $L(f)$ correspond to  $p < p^*$ and $p > p^*$. Another kind of 
  branching occurs at extrema of $H(p)$, if any, and the number of Lagrangians $L(f)$ on the 
  way from infinity to the center equals the number of monotonicity ranges of $f(p)$ \cite{k-NED}. 

  It might seem that the Hamiltonian' framework is not worse than the Lagrangian one, even
  though the latter is directly related to the least action principle. However, as shown in 
  \cite{k-NED}, at $p=p^*$ the electromagnetic field exhibits a singular behavior, well revealed 
  using the effective metric \cite{Nov-1, Nov-2} in which NED photons move along null geodesics.
  This metric is singular at extrema of $f(p)$, and the effective potential for geodesics exhibits 
  infinite wells at which NED photons are infinitely blueshifted \cite{k-NED, Nov-2} and can thus 
  create a curvature singularity by their back reaction on the metric. Thus any regular electric 
  solution not only fails to correspond to a fixed Lagrangian $L(f)$ but has other important 
  undesired features.

  Each choice of $A(r)$ thus gives rise to both electric and magnetic solutions related by FP duality,
  which connects the $F\mn$ and $P\mn$ frameworks \cite{k-NED}. Though, the choice of $A(r)$ 
  can be restricted by imposing the weak and dominant energy conditions (see, e.g., \cite{Bal-15}), 
  and individual features of numerous special solutions also deserve discussion.

\section{Dyonic configurations}

  Let us now assume that both $q_e$ and $q_m$ are nonzero. The difficulty with considering 
  \eqs \rf{SET} and \rf{A} is that $f(r)$ (or alternatively $p(r)$) is now not known explicitly.
  Using, for definiteness, the Lagrangian formulation of the theory, we have the expression
\beq                                     \label{f_r}
	 f(r) = \frac{2}{r^4} \biggl(q_m^2 - \frac{q_e^2}{L_f^2}\biggr). 
\eeq
  Comparing the expressions for $\cE(r)$ from \rf{SET} and from \rf{A}, we can write
  ($M' \equiv dM/dr$)
\beq    			      \label{M'}
           \Half L(f) + \frac{2q_e^2}{L_f r^4} = \frac{2M'(r)}{r^2} = \cE(r). 
\eeq

  To obtain a solution, we can specify one function, it is either $L(f)$ or, for example, $A(r)$.
  If $A(r)$ is known, or equivalently $M(r)$ or $\cE(r)$, then \eqn{f_r} 
  expresses $r$ in terms of  $f$, $L_f$ and the two charges, and substituting it into \eqn{M'}, we
  arrive at a highly nonlinear first-order differential equation with respect to $L(f)$. This 
  method does not seem fruitful since there is very little hope to solve such an equation, and 
  moreover, a choice of $A(r)$ or $M(r)$ is not physically evident: it is only clear (due to the 
  above-mentioned no-go theorem) that assuming the metric with a regular center, we cannot 
  obtain $L(f)$ with a Maxwell weak-field limit. 

  A better way is to specify the theory by choosing $L(f)$. Then \eqn{f_r} can be considered as either 
  (A) an equation (in general, transcendental) for the function $f(r)$ or (B) an expression of $r$ 
  as a function of $f$. 

  In case A, if $f(r)$ can be found explicitly, integration of \eqn{M'} (equivalent to \rf{A}) 
  gives the metric function $A(r)$.

  The scheme B leads to a solution in quadratures in terms of $f$ which can then be 
  chosen as a new radial coordinate. Indeed, assuming that $L(f)$ and $r(f)$ are known and monotonic, 
  so that $L_f \ne 0$ and $r_f \ne 0$, \eqn{M'} can be rewritten as 
\beq                                   \label{M_f}
               M_f = \frac{r^2 r_f}{2}\biggl[\frac{L}{2} + \frac{q_e^2}{L_f r^4}\biggr]
\eeq  
  (the subscript $f$, as before, denotes $d/df$). Since the r.h.s. of \rf{M_f} is known, 
  it is straightforward to find $A(r)$ and to pass on to the coordinate $f$ in the metric.
  This gives us {\it a general scheme of finding dyonic solutions} under the above conditions.   	`
  
  Examples of interest of using method B are yet to be found, therefore 
  let us restrict ourselves to considering two examples in which method A can work.

  The first example is trivial: the Maxwell theory, $L=f$, it is used to verify the method. Substituting 
  $L=f$ and $L_f=1$ to \eqn{M'}, we obtain $2M' = (q_e^2+q_m^2)/r^2$, whence 
  $2M(r) = 2m - (q_e^2+q_m^2)/r$ and 
\beq                    \label{RN}
                     A(r) = 1 - \frac{2m}{r} + \frac{q_e^2+q_m^2}{r^2}, \ \ \ m=\const,
\eeq 
  that is, the dyonic Reissner-Nordstr\"om solution, as should be the case. 

  In the second example we assume that \eqn{f_r} is linear with respect to $f$, which unambiguously
  leads to the truncated Born-Infeld Lagrangian,
\beq                  \label{BI}
	L(f) = b^2 \big(-1 + \sqrt{1+ 2 f/b^2}\big), \ \ \ b = \const
\eeq
  (it differs from the full Born-Infeld Lagrangian by the absence of the invariant $(^*F\mn F\MN)^2$).
      
  Indeed, for \eqn{f_r} to be linear in $f$, we have to assume $L_f^{-2} = c_1 f + c_2$ with 
  $c_{1,2} = \const$. Integrating this condition, we get $L = L_0 + (2/c_1)\sqrt{c_1f + c_2}$. 
  The requirement of a Maxwell behavior, $L \approx f$, at small $f$ leads to $c_2 =1$, $L_0=-2/c_1$.
  Then, denoting $2/c_1 = b^2$, we arrive at \eqn{BI}.

  With \rf{BI}, we obtain
\bearr              \label{f-BI}
            f(r) = \frac{2b^2 (q_m^2 - q_e^2)}{4q_e^2 + b^2 r^4},
\nnn                   \label{E-BI}
	   \cE(r) = -\frac{b^2}{2} + \biggl(\frac{b^2}{2} + \frac{2q_e^2}{r^4}\biggr)
			\sqrt{\frac{4q_m^2 + b^2 r^4}{4q_e^2 + b^2 r^4}}.
\ear  
   The simplest solution is obtained in the special case of a self-dual electromagnetic field, 
   $q_e^2 = q_m^2$, leading to $f=0$ and $L_f =1$, the same as for the Maxwell theory. This leads to
   $\cE = 2q_e^2/r^4$ and the dyonic Reissner-Nordstr\"om metric with $A(r)$ given by \rf{RN}.

   In the more general case, $q_e^2 \ne q_m^2$, $A(r)$ is a bulky expression in terms of 
   the Appel hypergeometric function $F_1$, not to be presented here. There are, however, 
   some features of the solution to be noticed. First of all, as expected, at large $r$ the quantities
   $f_e$. $f_m$ and the energy density $\cE$ decay as $r^{-4}$, and the solution is \asflat.
   At small $r$, from \rf{E-BI} it follows that $f(r)$ tends to a finite limit while 
   $\cE \approx 2 |q_e q_m|/r^4$. However, this relation does not hold in the cases of pure electric
   or magnetic solutions: in the electric case we also have $\cE \sim r^{-4}$ whereas in the 
   magnetic case $\cE  \sim r^{-2}$. This confirms that $r=0$ is a singular center in all dyonic 
   solutions and that the magnetic solution must also be singular since the function \rf{BI}
   does not tend to a finite limit as $f\to \infty$. However, in this theory, the energy density 
   in the pure magnetic solution blows up at the center slower than in all other cases,  so the 
   curvature singularity must be also milder. 

\section{Conclusion}

  The main results of this study are a scheme of finding dyonic solutions in NED coupled 
  to GR by quadratures for an arbitrary Lagrangian function $L(f)$ (up to some monotonicity
  restrictions) and a dyonic solution for the truncated Born-Infeld theory. A feature of interest 
  in this solution is the existence of a special case with a self-dual electromagnetic field and the 
  Reissner-Nordstr\"om metric.    

  We did not discuss here the question of horizons and global causal structure of the resulting 
  configurations. It is clear, however, that since the positive energy density $\cE(r)$ makes a 
  positive contribution to the metric function $A(r)$ (and even leads to a repulsive central 
  singularity like the Reissner-Nordstr\"om one) while the Schwarzschild mass $m$ appearing there as
  an integration constant can raise or lower the plot of $A(r)$ at intermediate $r$ (its influence
  at small $r$ is weaker than that of the electromagnetic field), the situation with horizons as 
  zeros of $A(r)$ in dyonic solutions is the same as is known in numerous electric and magnetic 
  ones: assuming fixed $q_e$ and $q_m$, solutions with small $m$ have no horizons, their central
  singularities are naked; at some critical value of $m$, depending on the charge values, a double
  (extremal) horizon appears, and at still larger $m$ there are two horizons. The space-time 
  causal structure is then quite similar to that of the Reissner-Nordstr\"om solution.

\subsection*{Acknowledgments}

  I thank Milena Skvortsova and Sergei Bolokhov for helpful discussions. 
  The work was partly performed within the framework of the Center FRPP 
  supported by MEPhI Academic Excellence Project (contract No. 02.a03.21.0005, 27.08.2013).
  The work was also funded by the RUDN University Program 5-100 and by RFBR grant 16-02-00602.


\end{document}